\documentclass[aps,showpacs,preprintnumbers,amsmath,amssymb]{revtex4}
\usepackage{graphicx}
\usepackage{epstopdf}
\usepackage{color}
\usepackage{subfig}
\usepackage{aasmacros}
\usepackage{amsmath,amssymb,amsfonts}
\usepackage{accents}

\begin{document}

    \baselineskip=0.8cm
    \title{\bf Signatures of Lorentz violation in the bright ring for Sgr A* images from radiatively inefficient accretion flows }

    \author{Cuiyu Zhao,
    Songbai Chen\footnote{Corresponding author: csb3752@hunnu.edu.cn},
    Jiliang Jing \footnote{jljing@hunnu.edu.cn}}
    \affiliation{Department of Physics, Institute of Interdisciplinary Studies, Hunan Research Center of the Basic Discipline for Quantum Effects and Quantum Technologies, Key Laboratory of Low Dimensional Quantum Structures
    and Quantum Control of Ministry of Education, Synergetic Innovation Center for Quantum Effects and Applications, Hunan
    Normal University,  Changsha, Hunan 410081, People's Republic of China}

    \begin{abstract}
    \baselineskip=0.6 cm
    \begin{center}
    {\bf Abstract}
    \end{center}
     
We have investigated the effects of Lorentz violation (LV) on the bright ring in Sgr A* images illuminated by the 230 GHz thermal synchrotron emission from 
 radiatively inefficient accretion flows around a rotating LV black hole within the low-energy Ho\v{r}ava gravity framework. Our results reveal that the LV parameter reduces the bright ring diameter yet increases its width, luminosity, azimuthal asymmetry and orientation angle. A higher spin parameter strengthens the LV-induced effects on the bright ring properties.
Increasing disk thickness reduces the ring diameter and enhances the LV parameter’s effects on this diameter. The ring width shows no systematic dependence on the disk thickness. These quantities of the bright ring display similar trends against the black hole spin and the LV parameter for the rotating LV black hole.
Using EHT observational data of Sgr A*, we find that, at fixed disk thickness, the parameter-space map illustrates that the bounded range of the LV parameter first broadens and then contracts with the growing black hole spin, and shifts toward smaller LV parameter values. In addition, the LV parameter narrows the bounded range of black hole spin:  negative LV parameter values shift this range to a higher spin, while  positive values shift it to a lower spin. Finally, we probe effects of the LV parameter  on the peak position value and the width of the primary image and the $n=1$ photon ring for the rotating LV black hole. The peak positions and their widths nearly always decrease with the LV parameter. The peak position differences for various LV parameters are more pronounced for a purely Keplerian accretion flow. In addition, the primary image and the $n=1$ photon ring produced by purely radially free-falling flows are broader than their counterparts generated by purely Keplerian flows. 
    \end{abstract}

    \maketitle
    \newpage

    \section{Introduction}
The Event Horizon Telescope (EHT) has recently delivered horizon-scale images of M87* and Sgr A*,  which yield direct visual evidence for the existence of astrophysical black holes. This landmark achievement opens new avenues for testing gravitational theories in the strong-field regimes \cite{EventHorizonTelescope:2019dse,EventHorizonTelescope:2019ths,EventHorizonTelescope:2019pgp,EventHorizonTelescope:2019ggy,EventHorizonTelescope:2022wkp,EventHorizonTelescope:2022wok,EventHorizonTelescope:2022exc,EventHorizonTelescope:2022urf,EventHorizonTelescope:2022xqj}. The brightness distributions within the black hole images encode extensive information about the near-horizon electromagnetic radiation, which can be exploited to probe the ambient matter configuration and the accretion dynamics
around black holes. Therefore, investigations into black hole images provide an invaluable pathway to unravel strong-field physics and  verify theories of gravity, which has motivated extensive efforts to explore the imaging signatures of various black hole systems \cite{Yin:2025rao,Zhang:2024jrw,Chen:2022scf,Qin:2023nog,Zhang:2022klr,Vagnozzi:2022moj,Wang:2023jop}.

The bright emission ring constitutes one of the key features in the EHT black hole images, arising primarily from strong gravitational lensing of synchrotron radiation generated by black hole accretion flows \cite{Desire_2025}. Such bright rings can be further decomposed into finer substructures including the primary image and a sequence of photon rings \cite{Johnson:2019ljv}. The primary image is formed by photons emitted from the disk that reach the observer along geodesics with no latitudinal turning points, which is commonly referred to as the direct image with $n = 0$. Photon rings correspond to higher-order images of the same emitting source, characterized by the number $n \geq 1 $ of latitudinal turning points. The primary image is highly sensitive to the
detailed properties of the accretion flow and the surrounding emission configurations. In contrast, photon rings are signatures associated with the strong-field lensing effects originating from the photon sphere, which exhibit only limited dependence on astrophysics in the accretion disk. In particular,  the position and size of the photon ring with $n\rightarrow\infty$ is entirely determined by the spacetime properties of the black hole and is independent of the properties of the accretion disk \cite{Johnson:2019ljv}. Since the photon ring with $n\rightarrow\infty$ is almost overlapped with the other subrings and its  brightness is obscured by the other lower-order images, the current observational precision does not allow us to detect such a ring in real astronomical observations. However, for the next-generation black hole image observing campaigns in the future, one of the primary scientific goals 
is to isolate the primary image and photon ring components within the observed images, thereby disentangling the signatures arising from spacetime geometry and accretion flows. It is therefore essential to theoretically investigate the signatures of the primary image and photon rings for various black holes.
Using the REx algorithm \cite{chael2019simulating}, the EHT Collaboration explored several characteristic quantities of the bright emission ring for Kerr black holes, such as the diameter, width, orientation and asymmetry \cite{EventHorizonTelescope:2019ths,EventHorizonTelescope:2019ggy,EventHorizonTelescope:2022exc,achour2025blackholephotonring}.

The bright ring features in the black hole images  depend on the surrounding magnetized accretion flows. Fully describing the dynamics of accretion flows  generally relies on high-precision general relativistic magnetohydrodynamics (GRMHD) simulations \cite{EventHorizonTelescope:2019ths,EventHorizonTelescope:2019pgp}. However, owing to their high computational overhead, systematic exploration over the large parameter space remains a formidable challenge. Moreover, for the current EHT targets M87* and Sgr A*, the
surrounding hot plasma is interpreted as belonging to a radiatively inefficient accretion flow (RIAF), a structure that is typically hot, geometrically thick, and optically thin at the observing frequency of 230 GHz. Therefore, the semi-analytic RIAF model capturing the features of magnetized accretion flows near black holes \cite{Yuan_2003,Pu_2018} has been adopted to probe horizon-scale images of various black holes \cite{Jiang:2023img,Yan:2025mlg,Pu:2016qak}.  

Lorentz invariance serves as a fundamental symmetry underlying general relativity and quantum theory. However, many candidate theories attempting to unify gravity and quantum theory (e.g., string theory and loop quantum gravity) predict possible LV signatures at the Planck scale. A general theoretical framework for studying LV is the Standard-Model Extension, within which extra fields are introduced to couple non-minimally to gravity. When these extra fields acquire a nonzero vacuum expectation value, they act as a fixed background field that spontaneously breaks Lorentz symmetry.  The vector field $B_{\mu}$  in the Einstein–bumblebee model \cite{Casana_2018,Bluhm:2007bd,Kostelecky:2005ic,Bertolami:2005bh} and the rank-two antisymmetric tensor $B_{\mu\nu}$  in Kalb–Ramond theory \cite{Kalb:1974yc,Yang:2023wtu,Altschul:2009ae,Lessa:2019bgi} fall into this class of fields.  Within the Einstein–bumblebee and Kalb–Ramond models, Lorentz-violating effects have been investigated with respect to black hole shadows, quasinormal modes, Hawking radiation, and gravitational lensing \cite{Jha:2024xtr,Horava:2009uw,Aharony:1999ti,Colladay:1998fq,Lin:2026ewo,Zeng:2025kyv,Xu:2025iwg,Kumar:2020hgm,Liu:2024lve,Kuang:2022xjp,Wang:2021irh,Liu:2019mls,Chen:2025ypx,AraujoFilho:2024ykw,Liu:2022dcn,Filho:2022yrk}.
These studies provide  valuable insights into the nature of spacetime and the fundamental principles governing physics.
Recently, Devecio\v{g}lu and Park adopt a concise two-step method to obtain an exact Kerr-type solution
within the low-energy regime of Ho\v{r}ava gravity \cite{Devecioglu:2024uyi}, which
is a four-dimensional theory exhibiting spontaneous Lorentz symmetry breaking. This rotating LV black hole solution is asymptotically flat at spatial infinity and the corresponding spacetime effects induced by LV  are principally exhibited via gravitational fields and frame-dragging. 
The shadow and polarization images of the rotating LV black hole illuminated by thin accretion disks have been respectively studied in \cite{Liu:2025lwj,Zhao:2025nwi}, which demonstrate that increasing the LV parameter yields a prominent leftward-tilting ``D"-shaped critical curve, whereas reducing the LV parameter gives rise to a more elliptical, untilted inner shadow. Furthermore, variations in the LV parameter modify the distribution of polarized intensity and polarization orientation, particularly in the vicinity of the critical curve. Since the accretion flows surrounding most astrophysical black holes correspond to geometrically thick RIAF disks, a further investigation
of the bright ring features of LV rotating black holes  illuminated by RIAFs
is particularly significant for revealing the nature of LV effects.
Along theses lines, we will  study LV signatures imprinted on the bright ring in Sgr A* images generated by RIAFs around a rotating LV black hole within the low-energy Ho\v{r}ava gravity framework.

This paper is organized as follows: In Sec. II, we briefly introduce the rotating LV black hole and the RIAF model. In Sec. III, we adopt the REx algorithm to extract the features of the bright ring from the black hole images generated by GRRT, and then investigate the influences of the LV parameter and the disk thickness on these images. In Sec. IV, we further analyse the effects of LV parameter on peak positions in the primary image and the $n=1$ photon ring $n=1$ for rotating LV black holes.
 Finally, we present a summary.

\section{MODELING RIAF DYNAMICS around rotating Lorentz-Violating black holes}
    \label{sec:2}
   The low-energy sector of non-projectable Ho\v{r}ava gravity admits 
   an exact solution of describing rotating Lorentz-Violating (LV) black holes, whose metric  in the Boyer-Lindquist coordinates can be expressed as \cite{Horava:2009uw,Devecioglu:2024uyi,Liu:2025lwj,Zhao:2025nwi} 
\begin{eqnarray}
\label{metric}
     ds^2 &=&-\bigg[1-\frac{2Mr}{\rho^2}-\frac{4l(l+2)M^2a^2r^2\sin^2\theta}{\rho^2\Sigma^2}\bigg]dt^2 +\frac{\rho^2}{\Delta}dr^2
     +\rho^2d\theta^2 \nonumber\\
&-&\frac{4(l+1)Mar\sin^2\theta}{\rho^2}dtd\varphi + \frac{\Sigma^2\sin^2\theta}{\rho^2}d\varphi^2,
\end{eqnarray}
with
\begin{eqnarray*}
\rho^2=r^2+a^2\cos^2\theta, \quad\quad\quad\Delta=r^2+a^2-2Mr,\quad\quad\quad\Sigma^2=\left(r^2+a^2\right)\rho^2+2Ma^2r\sin^2\theta.
\end{eqnarray*}
Here $M$ is the mass parameter of the black hole, and the parameter $l$ appearing in the $g_{tt}$ and $g_{t\varphi}$ components characterizes a nontrivial Lorentz symmetry–violating effect in spacetime and is referred to as the Lorentz-Violating parameter.
This solution can be returned to the Kerr solution when $l=0$ or to the Schwarzschild one when $a=0$. This means that the corrections involving the Lorentz-violating effects are only valid for rotating black holes. Actually, it can be explained by a fact that the corrected effects are entirely encoded in the shift function $N^{\varphi}$. From $ g^{rr}=\Delta/\rho^2=0 $, one can find that the outer and inner horizon radii $ r_{\pm} $ of the black hole are
\begin{equation}
r_\pm=M\pm\sqrt{M^2-a^2}, 
\end{equation}
which are consistent with those in the Kerr spacetime. Although the horizon locations remain unchanged, the
shadow shape \cite{Liu:2025lwj} and the observable thin-disk images \cite{Zhao:2025nwi} of the black hole 
are shown to differ markedly from those of the Kerr black hole. Motivated by these results,
our aim is to further investigate whether the bright ring in the observable thick disk images around this black hole exhibits significant variation as the LV parameter $l$  increases.

Since the hot plasmas surrounding the current observed targets of the EHT including M87* and Sgr A* are considered to be part of a RIAF, we here adopt a semi-analytical optically thin RIAF model \cite{Yuan_2003,Pu_2018} around the rotating Lorentz-Violating black hole to perform general relativistic ray-tracing numerical simulations \cite{Noble:2007zx,Moscibrodzka:2017lcu}.  In this semi-analytical RIAF model, the electron number density and temperature are assumed to follow power-law distribution with respect to the radius, whose forms can be respectively expressed as
\begin{align}
    n_{\rm e} &= n_{\rm e,0} \ r ^{-\delta} \exp \left[- \frac{1}{2} (H \tan \theta)^{-2} \right], \label{eq:ne} \\
    T_{\rm e} &= T_{\rm e,0} \ r ^{-\gamma}, \label{eq:Te}
\end{align}
where $\delta$ and $\gamma$ are the power-law indices for the  electron density and the  electron temperature respectively. From the observational constraints and previous studies of RIAFs \cite{Pu_2018,Yuan_2003,Broderick:2010kx,Yfantis:2024eab}, these two indices can be set to $\delta = 1.1$ and $\gamma = 0.84$, respectively. The parameter $H$ describes the geometric thickness of the disk and increasing the thickness makes the flow geometry approach a near-spherical configuration \cite{Pu_2018}.

The motion of flows within  realistic accretion disks cannot be fully described by either Keplerian rotation or free-fall solutions \cite{Narayan:2012yp}. Instead, the overall velocity may be a superposition of a radial flow component and a Keplerian rotational term \cite{Yuan:2022mkw,Begelman:2021ufo}. Therefore, in this semi-analytical RIAF thick disk model, the four-velocity and angular velocity of the accretion flow are modeled by interpolating between Keplerian motion and free-fall motion
\cite{Pu_2018,Pu:2016qak} 
    \begin{align}
        u^r = u^r_{\rm K} + \kappa_{\rm ff}(u^r_{\rm ff} - u^r_{\rm K}),  \quad\quad\quad\quad
        \Omega = \Omega_{\rm K} + (1-\kappa_{\rm K})(\Omega_{\rm ff} - \Omega_{\rm K}),\label{eq:Omega}
    \end{align} 
where $(u^r, \Omega)$ and $(u^r_{\rm ff}, \Omega_{\rm ff})$ denote the radial component of the four-velocity $u^{\mu}$ and the angular velocity $\Omega = u^\phi/u^t$ of flows moving along Keplerian and free-fall orbits, respectively. The coefficient $\kappa$ is the corresponding regulating  parameter.  The motion of flows corresponds to purely Keplerian motion as $(\kappa_{\rm ff},\kappa_{\rm K}) = (0,1)$ and to  free fall motion $(\kappa_{\rm ff},\kappa_{\rm K}) = (1,0)$.
For the flows in the sub-Keplerian state, the regulating parameters satisfy 
$0<\kappa_{\rm ff}=1-\kappa_{\rm K} <1$. Considering that the non-thermal emission is less dominant at the current observation band of 230 GHz,  the synchrotron emissivity in the disk is modeled by a relativistic thermal (Maxwell-J\"{u}ttner) electron distribution. In our simulations, a MAD-like configuration \cite{Chen_2022,Jiang:2023img} is adopted because the magnetically arrested disk (MAD) model is favored by the current EHT observations \cite{EventHorizonTelescope:2022xqj}. Moreover, the coefficients in the polarized radiative transfer process corresponding to emission, absorption and Faraday effects are calculated via the fitted formulas in \cite{Pandya:2016qfh}. The normalized electron number density $n_{e,0}$ and the electron temperature $T_{e,0}$ are set to $n_{e,0} \approx 10^7 \rm cm^{-3}$ and  $T_{e,0} \approx 10^{11} \rm K$, respectively \cite{Pu_2018}, which ensures that the observed flux of thermal synchrotron radiation matches the measured value of Sgr A* \cite{Yin:2025rao,Urso:2025gos,Saurabh:2025kwb,Pu:2016qak}.  The detail fiducial values in our simulation are listed in Table \ref{table:param}.
\begin{table}
    \centering
    \begin{tabular}{c c c}
        \hline
        Parameter & Value & Parameter Description\\
        \hline
        $M_{\rm BH}$ & $4.3\times10^6\rm{M}_\odot$ & Black hole mass \\
        $D_{s}$ &  $8.3 \times 10^3 \rm{pc}$ & Distance to the source\\
        $\theta_{i}$ & $30^\circ$ & Inclination angle\\
        $\delta$ & 1.1 & $n_{\rm e}$ power law index\\
        $\gamma$ & 0.84 & $T_{\rm e}$ power law index\\
        $\kappa_{\rm K}$ & $0.5$ & Keplerian parameter\\
        $\kappa_{\rm ff}$ & $0.5$ & Radial infall parameter\\
        $n_{\rm e,0}$ & $\approx10^7\,\rm{cm}^{-3}$ & Number density\\
        $T_{\rm e,0}$ & $\approx10^{11}\,\rm{K}$ & Temperature of electrons\\
        $\nu_{\rm obs}$ & $230$\,GHz & Observing frequency\\
        FOV & $200\,\mu\rm{as}$ & Field of view\\
        \hline
    \end{tabular}
    \caption{Fiducial parameters for Sgr A* and the surrounding RIAFs. }
    \label{table:param}
    \end{table}
 In Fig. \ref{fig:1}, we present several images of the rotating LV black hole based on the semi-analytical optically thin RIAF model \cite{Yuan_2003,Pu_2018}. The synthetic images were generated using the ipole GRRT code \cite{Noble:2007zx,Moscibrodzka:2017lcu} with adaptive ray tracing, and all images are normalized to a benchmark average flux density of 2.4 Jy. Null geodesics are integrated by directly solving the full geodesic ODE system numerically, without assuming the existence of the Carter constant. This ensures that the algorithm can also be applied to the modified LV spacetime metric (\ref{metric}). The top row shows the non-rotating case \(a=0\), where LV corrections vanish, providing a useful sanity check. As the black hole spin $a$ and the LV parameter $l$ grow towards positive values,  both the width and intensity of the bright ring in the black hole images are found to increase.  
\begin{figure}    
 \centering
\includegraphics[width=\textwidth]{fig_1.pdf}
\caption{Images of the rotating LV black hole for different values of the black hole spin $a$ and the LV parameter $l$ for a disk thickness of $H = 0.5$. The rows from top to bottom correspond to $a=0,\ 0.50,$ and $ 0.99$, respectively. The top row shows the non-rotating case $a=0$, where LV corrections vanish, providing a useful sanity check. The left, middle and right panels correspond to   $l=-0.99,\ 0,$ and $ 0.99$, respectively. Here we set $M=1$.}\label{fig:1}
\end{figure} 

\section{ Bright RING FEATURES in BLURRED IMAGES OF rotating LV BLACK HOLES}
    \label{sec:4}          
\begin{figure}   
\centering
\includegraphics[width=\textwidth]{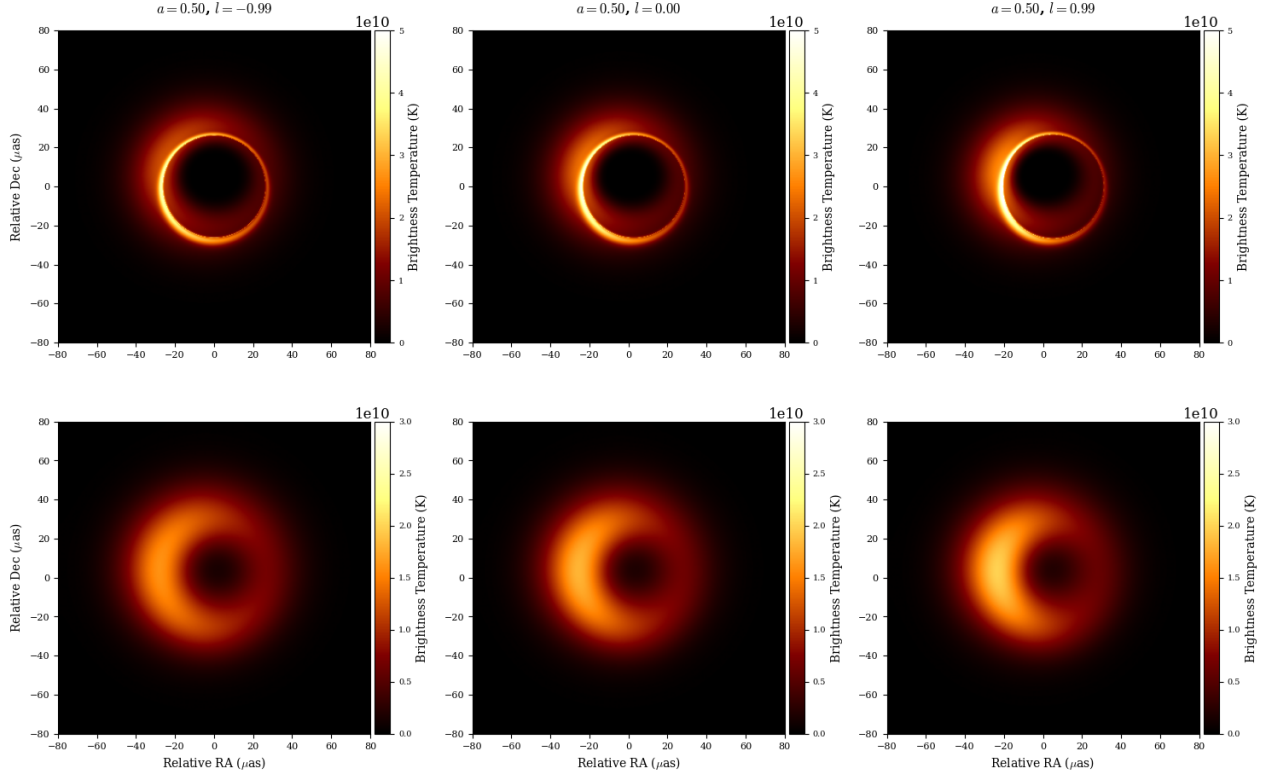}
\caption{Black hole images (in the top row) and the corresponding blurred images (in the bottom row), for an inclination angle of $30^\circ$, a disk thickness of $H = 0.5$ and a spin of $a=0.5$. The left, middle and right panels correspond to the cases with $l=-0.99,\ 0,\ 0.99$, respectively.} \label{fig:2}
\end{figure}

  Given the current limits of EHT observational resolution, extracting luminous ring features from GRRT-simulated blurred black hole images is essential for benchmarking theoretical models against the images actually captured by the EHT. In Fig. \ref{fig:2}, we present the blurred images of the rotating LV  black holes with a FWHM of 20 $\mu as$.
In order to extract the ring feature from the blurred images by the EHT REx algorithm \cite{EventHorizonTelescope:2019ths,chael2019simulating}, one must start from the center position $(x,y)$ of the candidate ring, sample the linearly interpolated image at equal intervals across azimuth angle $\theta$ of $0^{\circ}$ to $360^{\circ}$ and the radius $r$ of 0 to 50 $\mu as$, and then generate the transformed image $I(r,\theta,x,y)$. The function  $I(r,\theta,x,y)$ describes the brightness distribution in polar coordinates $(r,\theta)$  with the origin at the candidate center of the ring $(x,y)$. 
 The ring diameter $d$ is defined as twice $\bar{r}_{\rm pk}$ measured from the identified center $(x_0,y_0)$,  i.e., $d = 2\bar{r}_{\rm pk}(x_0,y_0)$. The quantity $\bar{r}_{\rm pk}(x,y)$ stands for the average of peak distances
 $r_{\rm pk}(\theta; x, y)$, where each $r_{\rm pk}(\theta; x, y)$ is defined as the distance at which the angular profile reaches its peak brightness
 \begin{align}
       r_{\rm pk}(\theta; x, y)={\rm argmax}_r[I(r,\theta,x,y)],\quad\quad\quad \bar{r}_{\rm pk}(x,y)= \langle r_{\rm pk}(\theta; x, y)\rangle_{\theta\in [0,2\pi]},
    \end{align}         
The ring center $(x_0, y_0)$ is identified by the position that minimizes the normalized radial peak
\begin{align}
        (x_0,y_0)=\rm argmin\bigg[\frac{\sigma_{\bar{r}(x,y)}}{\bar{r}_{pk}(x,y)}\bigg]_{(x,y)},
    \end{align}    
where $\sigma_{\bar{r}(x,y)}$ is the standard deviation of the $\bar{r}_{\rm pk}(\theta; x, y)$ values. 
The ring width $w$  is defined as the azimuthal average of the full width at half maximum (FWHM) of the radial profile
    \begin{align} 
        w=\text{FWHM}[I(r,\theta)-I_{\rm floor}].
    \end{align}
 Here the subtracted value is $I_{\rm floor}=\langle I(r=50\mu as,\theta)\rangle_{\theta}$,  which is introduced to mitigate measurement bias due to the non-zero intensity floor outside the ring in the resampled image $I(r,\theta)$. 
The ring orientation angle $\eta$, measured east of north, can be defined as
\begin{align}
\eta=\left\langle\operatorname{Arg}\left[\int_{0}^{2 \pi} I(\theta) e^{\mathrm{i} \theta} \mathrm{~d} \theta\right]\right\rangle_{r \in\left[r_{\mathrm{in}}, r_{\mathrm{out}}\right]}, \label{oritenangle}
\end{align}
where $r_{\mathrm{in}}=(d-w) / 2$ and $r_{\mathrm{out}}=(d+w) / 2$. With the REx algorithm, one can evaluate the independent variable of the angular mode $m=1$ for the angular profile at each radius. The total azimuthal angle $\eta$ is derived as the circular mean of these angles over the ring width. Note that in the REx algorithm, $\theta$ begins at the north ($0^\circ$) and increases counterclockwise toward the east ($90^\circ$). 

\begin{figure}
        \centering
        \includegraphics[width=\textwidth]{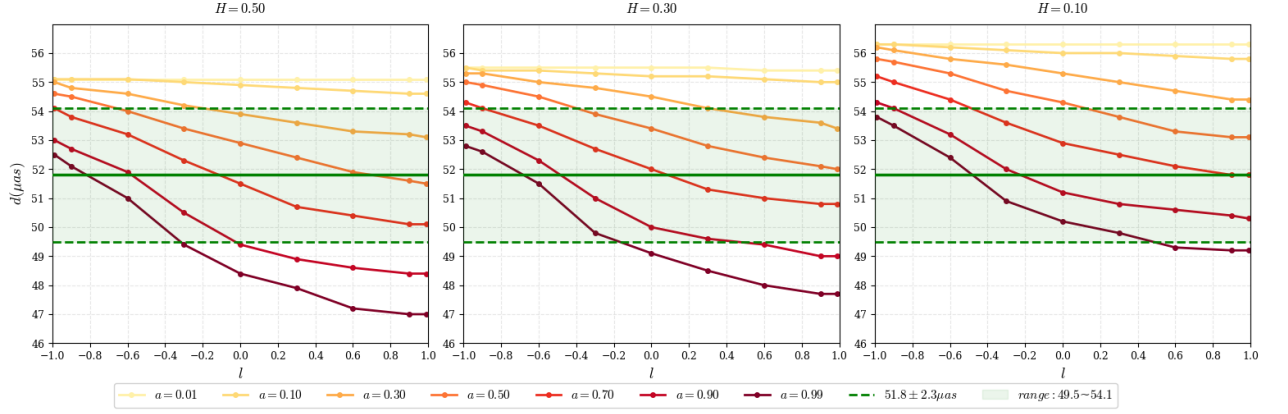}
        \caption{The $d$-$l$ relationship for the simulated blurred LV black hole images for different thickness parameters $H$ and spin parameters $a$.  The left, middle, and right panels correspond to the thickness parameters $H = 0.5,\ 0.3,\ 0.1$, respectively.}
        \label{fig:4}
    \end{figure}

The degree of azimuthal asymmetry $A$ in the ring can be defined as the normalized amplitude of the first angular mode over the radial range from $r_{\mathrm{in}}$ to $r_{\mathrm{out}}$
\begin{align}
A = \left\langle \frac {\left| \int_ {0} ^ {2 \pi} I (\theta) e ^ {\mathrm {i} \theta} \mathrm {d} \theta \right|}{\int_ {0} ^ {2 \pi} I (\theta) \mathrm {d} \theta} \right\rangle_ {r \in \left[ r _ {\mathrm {i n}}, r _ {\mathrm {o u t}} \right]}.
\end{align}
The value of the asymmetry  $A$  lies in the interval $[0,1]$, with $A=0$ indicating perfect azimuthal symmetry and $A=1$ corresponding to the extreme case where all flux density is confined to a single orientation angle. The brightness asymmetry of the ring in the black hole image can be described by the brightness ratio 
\begin{align}
{I}_{r}=\frac{{I}_{2}}{{I}_{1}},
\end{align}
where $I_{2}$ is the total brightness within the half-circle centered on the orientation angle $\eta$, while $I_{1}$ is the total brightness within the half-circle centered on $\eta+ 180^{\circ}$. This ratio quantifies the brightness asymmetry in the black hole image.
\begin{figure}
        \centering
        \includegraphics[width=\textwidth]{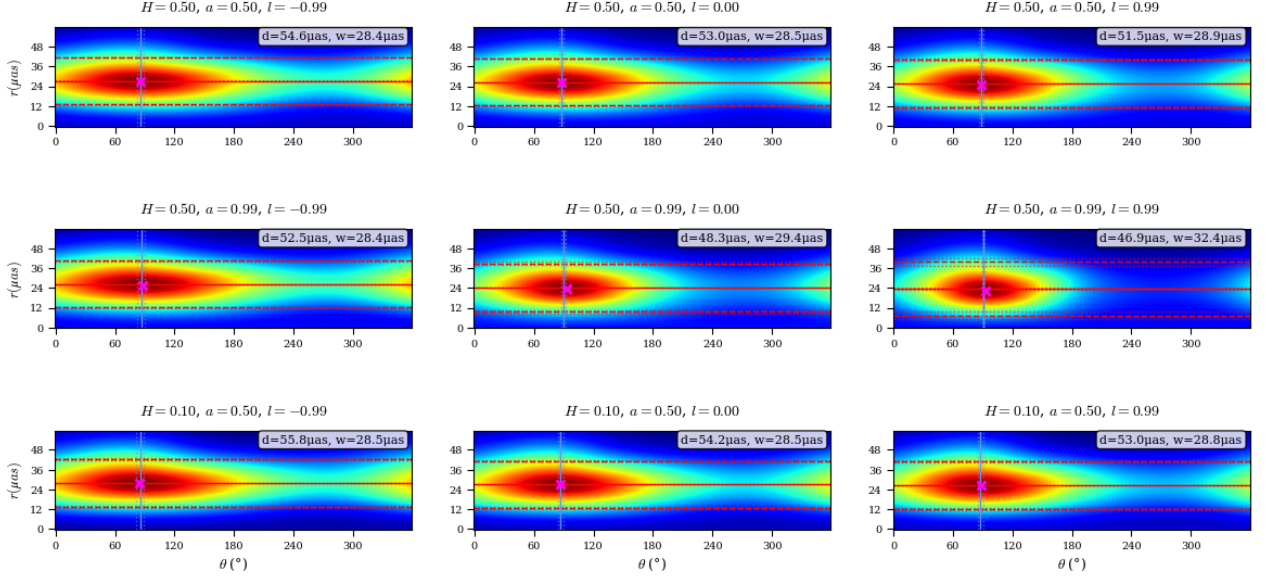}
        \caption{Unwrapped ring profiles of the simulated blurred LV black hole images for different values of the black hole spin $a$ and the disk thickness $H$. The left, middle and right panels correspond to $l = -0.99,\ 0,$ and $ 0.99$, respectively.}
        \label{fig:3}
\end{figure}    
Employing this REx algorithm, in Fig. \ref{fig:3}, we present the corresponding unwrapped ring profiles of the simulated blurred LV black hole images for different values of the spin parameter $a$, the LV parameter $l$ and the disk thickness $H$. As the LV parameter $l$ increases, the ring diameter $d$ decreases monotonically. 
The variation of the ring diameter $d$ with the LV parameter $l$ resembles its dependence on the black hole spin parameter $a$, which is also shown in Fig.\ref{fig:4}. As $H$ increases, the ring diameter $d$ also decreases. Moreover, we find that an increase in the thickness parameter $H$ enhances the impact of the LV parameter $l$ on the ring diameter $d$. The ring width $w$ increases with both the LV parameter $l$ and the black hole spin parameter $a$, but it exhibits  no systematic dependence on $H$. 
The dependence of these quantities on $l$ show behavior analogous to that on $a$. This can be explained by the fact that the LV parameter $l$ often enters the metric functions in the form of a product with the spin parameter $a$.

 According to the observational data from the EHT \cite{EventHorizonTelescope:2022wkp,EventHorizonTelescope:2022xqj,EventHorizonTelescope:2022exc,EventHorizonTelescope:2022urf,Chael:2018oym}, the diameter of the bright ring in the image of the supermassive black hole Sgr A* is $d_{\rm ring}=51.8\pm 2.3~\mu{\rm as}$.  In Fig. \ref{fig:4}, we present the bounded parameter range in the $l$-$d$ plane.
The solid and dashed green lines  respectively correspond to $d=51.8\mu as $ and its $1\sigma$ uncertainty range (i.e., $\pm 2.3\mu as$). For a fixed disk thickness $H$, the   bounded range of the LV parameter $l$ exhibits an initial increase followed by a decrease as the black hole spin $a$ increases. Meanwhile, the  bounded  range shifts toward lower $l$ with increasing $a$. Furthermore, these effects are enhanced as the disk thickness $H$ increases. From Fig. \ref{fig:4}, we also find that the presence of the LV parameter $l$ reduces the  bounded  range of the black hole spin $a$. Negative values of $l$ shift the  bounded  range toward higher spin, whereas positive values of  $l$ shifts it toward lower spin. Comparing the bounded regions from RIAF+REx with the vacuum-shadow bounds of Ref.\cite{Liu:2025lwj}, we find that the bounded range of the LV parameter $l$ obtained from  the two methods exhibits trends consistent with those found for the black hole spin $a$, featuring an initial increase followed by a decrease. However, there are some notable differences. In the low-spin regime, at $a=0.1$, their bounded range of the LV parameter corresponds to $l>0.47$, while our analysis shows that no $l$ values fall within our bounded parameter space. In the higher-spin regime, at $a=0.99$, their bound is $l<0.72$. In our results,  positive values of $l$  lie outside the bounded parameter space for $H=0.5$ and $0.3$, while $l<0.5$ is covered for $H=0.1$. This suggests that our derived bounded regions yield somewhat narrower ranges in this parameter-space comparison. These discrepancies may stem from several factors. First, we adopt RIAFs as the light source for Sgr A*, which could better capture the physical reality. Second, our analysis employs images blurred with a $20\mu\text{as}$ kernel and the bright-ring diameter rather than the critical curve, which could provide a better match with the actual observed quantities of the EHT. Moreover, we choose  the observer's  inclination $30^\circ$ rather than $90^\circ$ as adopted in [51].
\begin{figure}
        \centering
        \includegraphics[width=\textwidth]{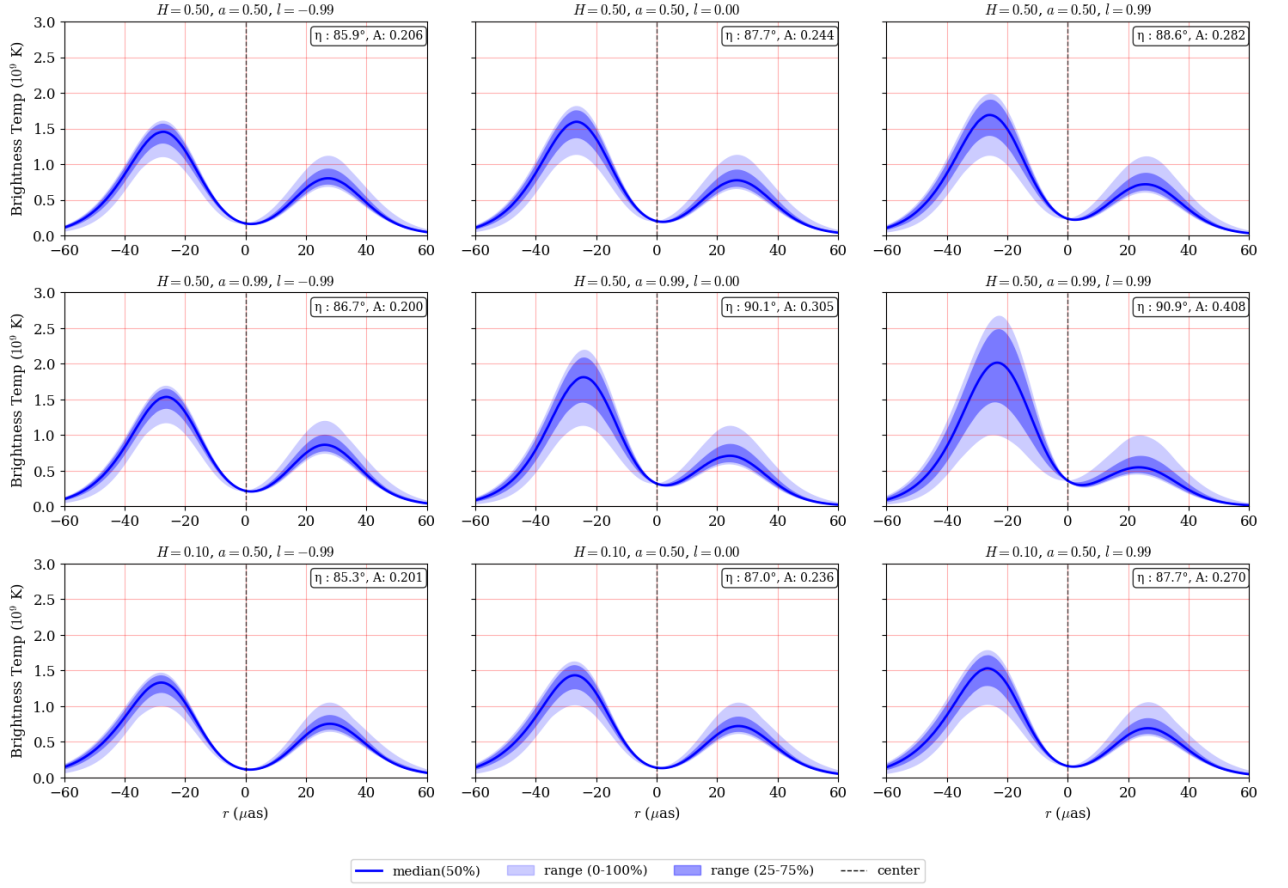}
        \caption{ The one-dimensional radial brightness profiles corresponding to the same parameters as in Fig. \ref{fig:3}.}
        \label{fig:5}
    \end{figure}
\begin{figure}
        \centering
        \includegraphics[width=\textwidth]{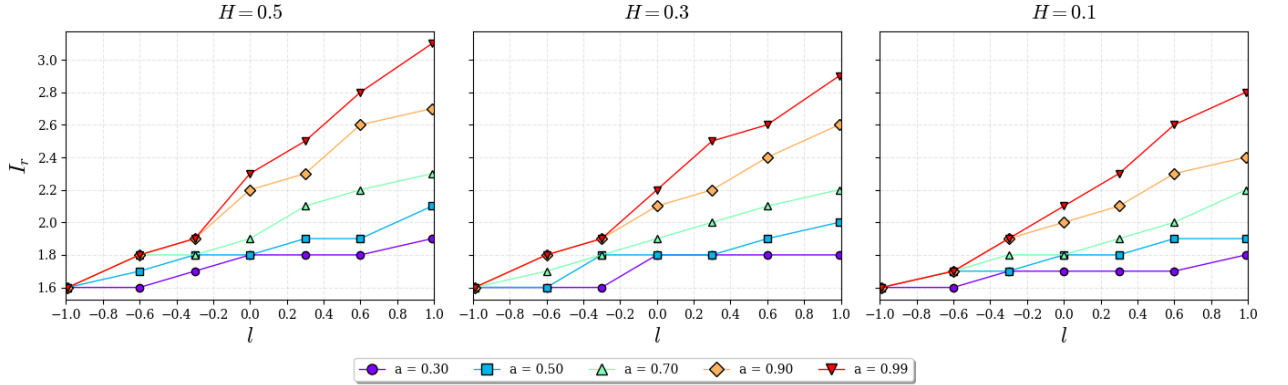}
        \caption{ The relationship between the brightness asymmetry $I_{r}$ and the LV parameter $l$ for the simulated blurred LV black hole images with different thickness parameters $H$ and black hole spin parameters $a$.}
        \label{fig:6}
    \end{figure}

In Fig.\ref{fig:5}, we present the orientation angle $\eta$ and the azimuthal asymmetry $A$ of the bright ring from the radial brightness profiles corresponding to the same parameters as in Fig.\ref{fig:3}. Radial profiles for the semicircle mirrored across the line at position angle $\eta$ are assigned negative $r$, and those for the opposing semicircle centred at position angle $\eta+180^{\circ}$ are assigned positive $r$. The solid curves represent the median profile for each semicircle. The dark shaded band illustrates the $25\% - 75\%$ range, while the light shaded band covers all profiles from the fiducial images. The orientation angle $\eta$ and the azimuthal asymmetry $A$ of the bright ring are listed in the upper right corner in each panel. From Fig.\ref{fig:5}, it is clearly seen that the radial brightness profiles of the black hole image are asymmetric rather than perfect azimuthal symmetry. As the LV parameter $l$ increases from negative to positive values, the orientation angle $\eta$, the azimuth asymmetry $A$, and the width of the brightness distribution range increase simultaneously. In particular, the width of the brightness distribution within the $25\% - 75\%$ range increases for the semicircle containing the orientation angle $\eta$, whereas the contrast inside the semicircle on the $\eta+180^{\circ}$ side slightly decreases.
Similarly, an increase in the spin parameter $a$ exacerbates this trend. In particular, when both the LV parameter $l$ and the spin parameter $a$ rise to 0.99, the azimuthal asymmetry $A$ and the extent of the brightness distribution grow considerably, leading to stronger deviation from perfect azimuthal symmetry. The brightness distributions in the semicircle containing the orientation angle $\eta$ and the $\eta+180^{\circ}$ semicircle become more distinct. Furthermore, we find that an increase in the disk thickness parameter $H$ also causes the orientation angle $\eta$ and the azimuthal asymmetry $A$ to increase.

In Fig.\ref{fig:6}, we also present the brightness asymmetry $I_r$ of the radial brightness profiles in the LV black hole images. It can be seen  that  $I_r$ is always greater than 1, which means that the brightness within the semicircle related to $\eta$  is always greater than that within the semicircle corresponded to $\eta+180^{\circ}$.
As the LV parameter increases from negative to positive values, the bright asymmetry $I_r$ also increases.  Similarly, the spin parameter together with the thickness parameter contributes to an increase in the brightness asymmetry $I_r$.
These results deepen our understanding of the characteristic features of LV black hole images and offer a potential route to probe Lorentz violation through future high-resolution observations of such images.

\section{Peak positions in the primary image and the $n=1$ photon ring for rotating LV black holes}
\label{sec:disc}
 
 In general, black hole images  can be decomposed into three components: the inner shadow, the primary image, and the photon rings \cite{Urso:2025gos,Paugnat:2022qzy,Khan:2025qsy,Bardeen:1973tla,Johnson:2019ljv,Gralla:2019xty,Chael:2021rjo}.
The inner shadow of a black hole refers to the dim  region in the black hole image. The primary image is formed by photons emitted from the disk that reach the observer along geodesics without winding  around the black hole; this component  is commonly referred to as the direct image with $n=0$. Photon rings correspond to higher-order images of the same emitting material, characterized by  the number $n \geq 1$ of windings around the black hole.
The primary image is highly sensitive to the detailed properties of the accretion flow and the surrounding emission configurations. In contrast, photon rings are signatures associated with the strong-field lensing effects originating from the photon sphere, which exhibit only limited dependence on astrophysics in accretion disks. Therefore, it is  essential to study the direct image and the photon rings separately. For a geometrically thin accretion disk located in the equatorial plane, it is possible to
decompose the simulated image into different $n$ orders by keeping track of the equatorial crossings. However, this decomposition criterion is not well suited to geometrically thick accretion disks such as RIAF disks, because the emission is not confined to the equatorial plane. It is well known that geodesics passing close to the black hole may undergo latitudinal oscillations due to strong lensing. For a geometrically thick accretion disk, a single light ray may traverse different parts of the emitter sequentially, and each emission event $x^{\mu}$ along a non-equatorial geodesic can be assigned an orbit number $n$, which counts the number of latitudinal turning points between the event and the observer, evaluated in the region where the geodesic crosses the midplane. For a geometrically thin disk, $n$ can equivalently be taken as the
number of windings (half-orbits) around the black hole; the two definitions coincide
for near-equatorial rays, which is why the equatorial-crossing criterion suffices in
that limit. This quantity can be directly tracked and stored during backward geodesic integration. The $n$-th subring resides inside the $(n-1)$-th, and different segments of a single geodesic that experiences $m$ latitudinal oscillations contribute to the first $m+1$ subrings. The primary image with $n = 0$ can be obtained by integrating the specific intensity along the geodesic segments with no latitudinal turning points, whereas the $n=1$ image is the $I_{\nu}$ map accumulated along the geodesic segments with exactly one latitudinal turning point.  Here, we focus on  these cases $n=0$ and $n=1$ and study the corresponding effects originating from the LV parameter $l$, the black hole spin $a$ and the disk thickness $H$.
  \begin{figure}
        \centering
        \includegraphics[width=\textwidth]{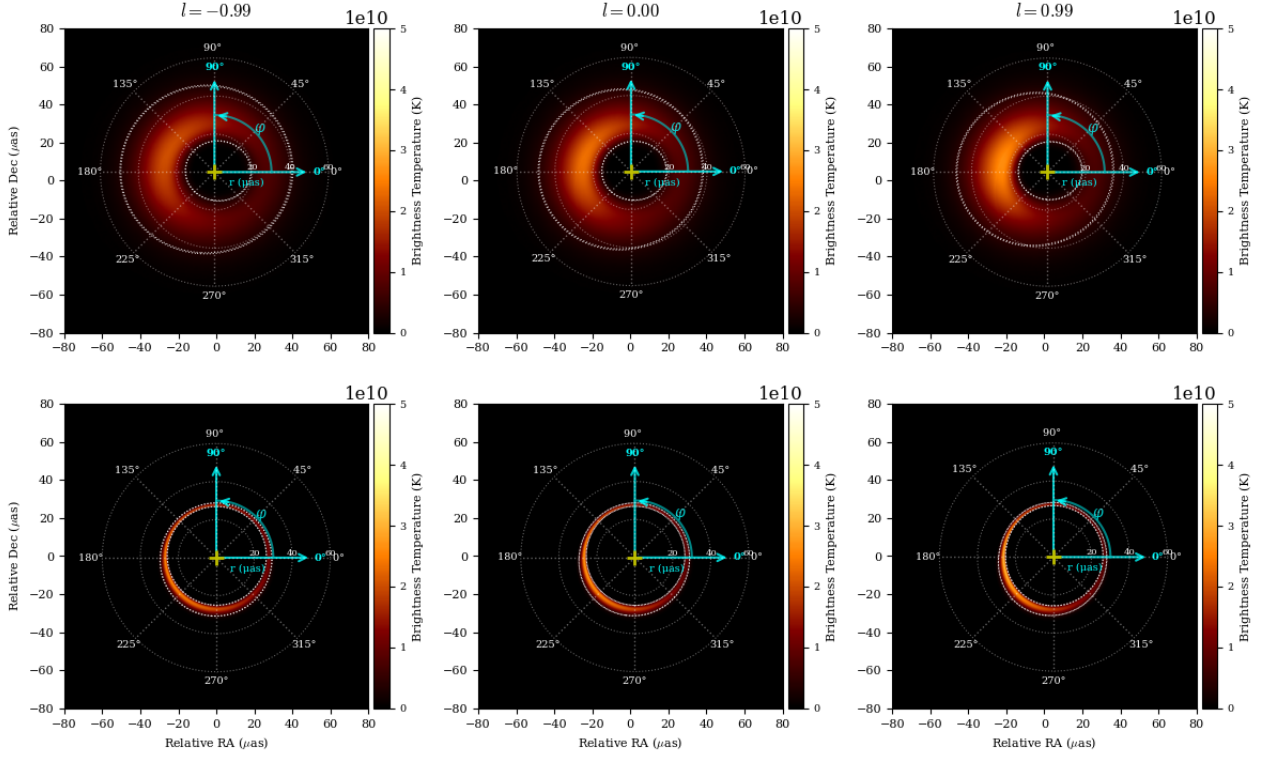}
        \caption{Intensity profiles for the direct image and the $n=1$ photon ring of the rotating LV black hole obtained from intensity maps with fixed $H=0.5$ and $ a=0.5$. The left, middle and right panels correspond to cases with $l = -0.99$, $0$, and $0.99$, respectively. The upper and bottom rows correspond to the $n=0$ direct image and the $n=1$ photon ring, respectively. Here, $\varphi$ denotes the azimuthal angle measured counterclockwise from east.}
        \label{fig:7}
    \end{figure}
\begin{figure}
        \centering
        \includegraphics[width=\textwidth]{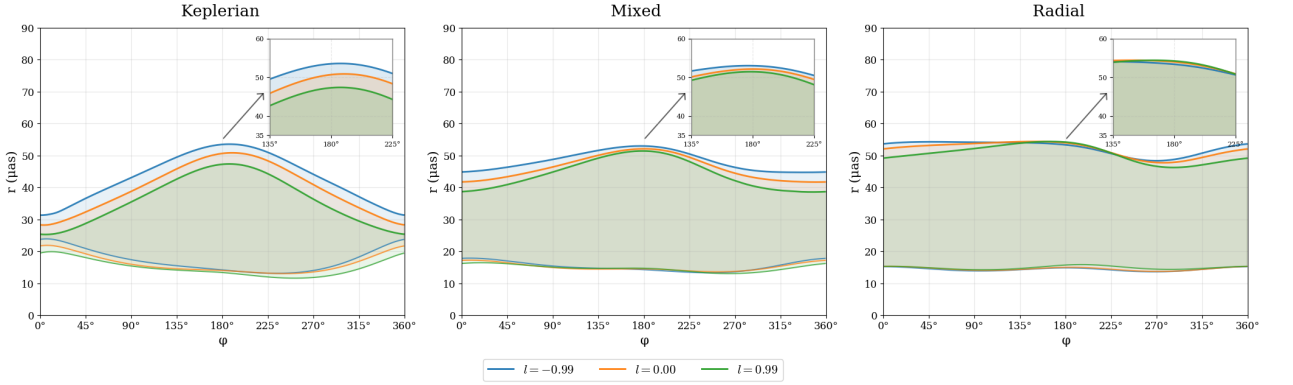}
        \caption{Peak positions of the $n = 0$ direct image  of the LV black hole with fixed $H=0.5$ and $ a=0.5$. Each column corresponds to a different velocity: Keplerian, mixed or radial, from left to right. The blue, orange and green lines represent the cases with $l = -0.99$, $0$, and  $0.99$, respectively.}
        \label{fig:8}
\end{figure}

One-dimensional intensity profiles, as an essential quantitative analysis approach for the primary image and  the photon ring of black holes, are expressed in terms of the impact parameter $b$ and can be extracted from the observed intensity of black hole images along a given azimuthal angle $\varphi$ \cite{Urso:2025gos}. Each profile contains a prominent intensity maximum termed the peak, and the peak location can used to identify the impact parameter associated with the maximum observed brightness along the given angular direction. The width of the intensity profile can be determined as the difference  between the radius of the maximum intensity peak and the radius of the innermost non-zero intensity isophote, i.e., the contour closest to the black hole image center.
From Fig. \ref{fig:7}, one can find that changing the LV parameters does not significantly alter the contours of the $n=0$ direct image and the $n=1$ photon ring. 
\begin{figure}
        \centering
        \includegraphics[width=\textwidth]{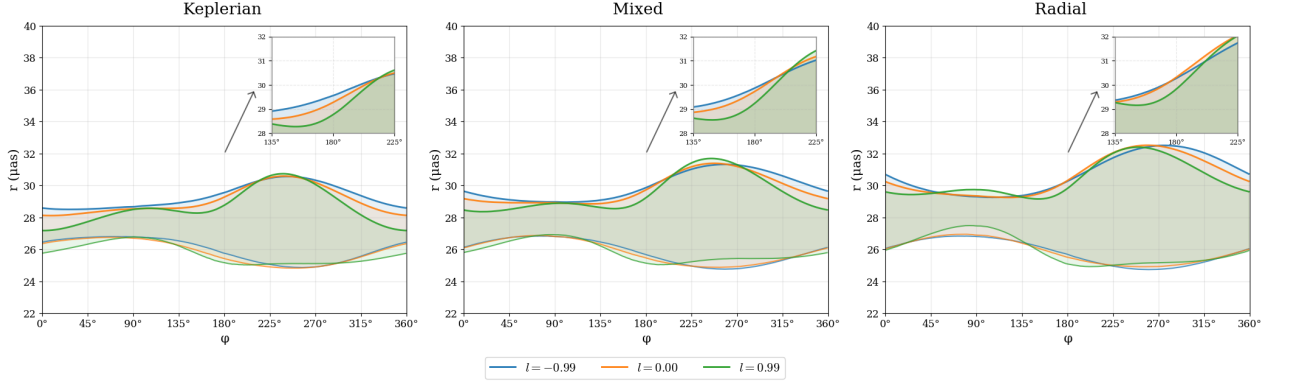}
        \caption{Peak positions of the $n = 1$ photon ring of the LV black hole with fixed $H=0.5$ and $ a=0.5$. Each column corresponds to a different velocity: Keplerian, mixed or radial, from left to right. The blue, orange and green lines represent the cases with $l = -0.99$, $0$, and  $0.99$, respectively.}
        \label{fig:9}
\end{figure}
In Figs. \ref{fig:8}-\ref{fig:9}, we present the effects of the LV parameter $l$ on the peak position value and  width of the intensity profiles of the rotating LV black hole for three type of accretion flows, namely, Keplerian, mixed or purely radial free-falling flows. 
For the $n = 0$ direct image and the $n = 1$  photon ring, we find that the peak position and width of the black hole image nearly always decrease as the LV parameter  $l$  increases. Moreover, for the $n = 0$ direct image, the peak position differences across various values of  $l$  are more pronounced for the Keplerian accretion flow, whereas the peak positions are more tightly clustered for the other two flow models. Furthermore, the primary image and the  $n = 1$  photon ring arising from radially free-falling flows are broader than their counterparts generated by Keplerian flows.
Although the $n=1$ photon ring is not accessible at current EHT resolution, our analysis of these fine structures in the primary image and the photon ring induced by the LV parameter $l$ could provide theoretical forecasts for probing LV effects in future high-resolution ngEHT/space-VLBI observations of black hole images.

\section{Summary}
\label{sec:6}
We have investigated the effects of Lorentz symmetry violation on the bright ring in Sgr A* images illuminated by the 230 GHz thermal synchrotron emission from
 RIAFs. Our results reveal that both the width and intensity of the bright ring in the black hole images increase with  the LV parameter $l$ and the black hole spin $a$, while the ring diameter $d$ decreases monotonically. As $H$ increases, the ring diameter $d$ decreases  and the effect of the LV parameter $l$ on the ring diameter $d$ is enhanced.
 Moreover, the ring width $w$ displays no systematic dependence on the disk thickness $H$. The dependence of these quantities on $l$ show behavior analogous to that on $a$, which can be explained by the fact that the LV parameter $l$ often enters the metric functions in the form of a product with the spin parameter $a$.
 We also probe the effects of the LV parameter $l$ on the orientation angle $\eta$, the azimuth asymmetry $A$ and the bright asymmetry $I_r$. All three quantities increase synchronously as the LV parameter $l$ increases. Furthermore, the spin and thickness parameters together enhance the orientation angle and both asymmetries. 

With the observational data of the supermassive black hole Sgr A* from EHT, we also present the bounded parameter range in the $l$-$d$ plane. For a fixed disk thickness $H$, the  bounded range of the LV parameter $l$ exhibits an initial increase followed by a decrease as the black hole spin $a$ increases. Meanwhile, the bounded range shifts toward lower $l$ with increasing $a$. Furthermore, these effects are enhanced as the disk thickness $H$ increases. We also find that the presence of the LV parameter $l$ reduces the bounded range of the black hole spin $a$. Negative values of $l$ shift the bounded range toward higher spin, whereas positive values of  $l$ shifts it toward lower spin.

Finally, we probe the effects of the LV parameter $l$ on the peak position value and the width of the primary image and the $n=1$ photon ring of the rotating LV black hole. Their peak positions and widths nearly always decrease as the LV parameter $l$ increases. The peak position differences for various values of $l$ are more pronounced for the Keplerian accretion flow. In addition, the primary image and the $n=1$ photon ring produced by radially free-falling flows are broader than their counterparts generated by Keplerian flows. These results deepen our understanding of the characteristic features of LV black hole images and offer  theoretical forecasts for probing  Lorentz violation in future high-resolution ngEHT/space-VLBI observations of black hole images.

\vspace{1cm}

\begin{acknowledgments}

This work was supported by the National Natural Science Foundation of China under Grant No.12675066, 12275078, 11875026, 12035005, 2020YFC2201400, the innovative research group of Hunan Province under Grant No. 2024JJ1006, and the 2026 Open Fund for Large-Scale Instrument Testing of Hunan Normal University Grant No. 26CSY151.
\end{acknowledgments}

\bibliography{LVBH}

\end{document}